\documentclass[namedreferences]{kluwer}   
\usepackage{graphicx}
\begin{document}

\begin{article}
\begin{opening}         
\title{Chemical Abundances in our Galaxy and Other \\
Galaxies Derived from H~{\sc{ii}} Regions} 

\author{Manuel \surname{Peimbert}}
\author{Leticia \surname{Carigi}}
\author{Antonio \surname{Peimbert}}  
\institute{Instituto de Astronom\'{\i}a, Universidad Nacional Aut\'onoma de
M\'exico \email{peimbert,carigi,antonio@astroscu.unam.mx}}

\runningauthor{Peimbert, Carigi, \& Peimbert}
\runningtitle{Abundances Derived from H~{\sc{ii}} Regions}


\begin{abstract}
We discuss the accuracy of the abundance determinations of H~{\sc{ii}} regions
in our Galaxy and other galaxies. We focus on the main observational
con\-straints derived from abundance determinations that have implications for
models of galactic chemical evolution: a)~the helium to hydrogen abundance
ratio, He/H; b)~the oxygen to hydrogen abundance ratio, O/H; c)~the carbon to
oxygen abundance ratio, C/O; d)~the helium to oxygen and helium to heavy
elements abundance ratios, $\Delta Y/\Delta O$ and $\Delta Y/\Delta Z$; and
e)~the primordial helium abundance, $Y_p$.
\end{abstract}

\end{opening}           

\section{Introduction}

The H~{\sc{ii}} region abundances provide important observational con\-straints
to test models of stellar evolution, models of the chemical evolution of
galaxies, and models of the evolution of the universe as a whole.

Reviews on the abundances of Galactic and extragalactic H~{\sc{ii}} regions
have been presented before (e. g. \opencite{pei93z}, \citeyear{pei99};
\opencite{gar99z}) \opencite{hen99}.

\section{Temperature differences and abundance determinations}

{From} emission line spectra of Galactic and extragalactic H~{\sc{ii}} regions
it is possible to determine abundance ratios, good quality spectra usually
permit to derive electron temperatures like $T_e$(O~{\sc{iii}}) and
$T_e$(N~{\sc{ii}}) from the [O~{\sc{iii}}] 4363/5007 and [N~{\sc{ii}}]
5755/6584 line ratios.

Often only $T_e$(O~{\sc{iii}}) is available and to estimate the temperature of
the regions of low degree of ionization photoionization models have been
used. For example, to determine $T_e$(O~{\sc{ii}}) \inlinecite{izo97},
\inlinecite{izo98}, and \inlinecite{deh00} have used the models by
\inlinecite{sta90} that provide a $T_e$(O~{\sc{iii}}) versus $T_e$(O~{\sc{ii}})
relationship. Photoionization models have been used also to estimate the
ionization correction factor for those elements that have not been observed in
all the stages of ionization present in the nebulae.

There are other methods to determine the electron temperature, like: the ratio
of the Balmer continuum to a Balmer line, the ratio of He~{\sc{i}}/H~{\sc{i}}
lines, and the ratio of recombination lines of C~{\sc{ii}} and O~{\sc{ii}} to
forbidden lines of [C~{\sc{iii}}] and [O~{\sc{iii}}]; these other methods
usually yield electron temperatures substantially smaller than those provided
by $T_e$(O~{\sc{iii}}) (e.g. \opencite{pei67}; \opencite{pei93y},
\citeyear{pei95y}; \opencite{liu00}). Several explanations have been proposed
for the $T_e$ differences, the main ones are: temperature variations, density
variations, and chemical inhomogeneities (e.g. \opencite{pei67},
\citeyear{pei71}; \opencite{tor90}; \opencite{vie94}; \opencite{liu00}). Recent
discussions on this subject have been presented by \inlinecite{pei95z},
\citeauthor{sta96} (\citeyear{sta96,sta98,sta00}), and \inlinecite{liu00}.

Detailed photoionization models of I~Zw~18, NGC~2363, and NGC~346 yield
$T_e$(O~{\sc{iii}}) values from 10\% to 15\% smaller than observed, probably
indicating that there are additional heating sources not considered by the
models, like the deposition of mechanical energy (\opencite{sta99};
\opencite{lur99}; \opencite{rel00}).

The differences in $T_e$ should be taken into account to derive accurate
abundance ratios.

\section{He/H}

To derive accurate He/H abundance ratios it is necessary to use accurate
temperatures and densities. Due to the collisional contribution to the
He~{\sc{i}} line intensities the higher the density the lower the computed He/H
ratio. Due to the density variations present in H~{\sc{ii}} regions it is well
known that $N_e$(rms) is smaller than the densities derived from the ratio of
two forbidden lines, in general it can be shown that $N_e^2{\rm (local)} >
N_e^2{\rm (rms)} = \epsilon N_e^2{\rm (local)}$, where $\epsilon$ is the
filling factor and for giant extragalactic H~{\sc{ii}} regions is typically of
the order of 0.01. Consequently the use of $N_e$(rms) yields a higher limit to
the He/H ratio.

For some objects only the $N_e$ derived from the [S~{\sc{ii}}] lines is
available and therefore the $N_e$(S~{\sc{ii}}) density has been used to
determine the He/H ratio.  \inlinecite{izo94} pointed out that the region where
the [S~{\sc{ii}}] lines originate is not representative of the region where the
He~{\sc{i}} lines originate, and proposed to determine self consistently the
density from five of the best observed He~{\sc{i}} lines, adopting
$T_e$(O~{\sc{iii}}) as the representative temperature for the regions where the
He~{\sc{i}} and H~{\sc{i}} lines originate.

Most He/H determinations have been made by adopting
$T_e$(O~{\sc{iii}}). $T_e$(O~{\sc{iii}}) provides us with an upper limit to
$T_e$(He~{\sc{ii}}) for the two following reasons: a) for metal poor
H~{\sc{ii}} regions $T_e$(O~{\sc{iii}}$) > T_e$(O~{\sc{ii}}), and the
He~{\sc{i}} lines originate both in the O~{\sc{iii}} and the O~{\sc{ii}} zones,
b) even if the [O~{\sc{iii}}] and the He~{\sc{i}} lines originate in the same
zone, in the presence of temperature variations it can be shown that
$T_e$(O~{\sc{iii}}$) > T_e$(He~{\sc{ii}}).

\inlinecite{pei00a} from nine He~{\sc{i}} lines of NGC~346, the brightest
H~{\sc{ii}} region in the Small Magellanic Cloud, derived self-consistently
$N_e$(He~{\sc{ii}}), He/H, and $T_e$(He~{\sc{ii}}). They derived a
$T_e$(He~{\sc{ii}}) value 9\% smaller than $T_e$(O~{\sc{iii}}). The maximum
likelihood method implies that the lower the temperature the higher the density
and the lower the derived He/H ratio, this is a systematic effect and implies
that the He/H ratios derived from $T_e$(O~{\sc{iii}}) are upper limits to the
real He/H value. {From} photoionization models of giant H~{\sc{ii}} regions
based on CLOUDY \cite{fer96} it is found that $T_e$(He~{\sc{ii}}) is from 3\%
to 12\% smaller than $T_e$(O~{\sc{iii}}) \cite{pei00c}.

\section{O/H}

The abundances of the Sun and the Orion nebula have been used as probes of
Galactic chemical evolution and as standards for stars and gaseous nebulae of
the solar vicinity. Therefore it is important to compare them since they have
been derived using different methods.

A decade ago the O/H difference between the Sun and the Orion nebula in the
literature amounted to 0.44dex, at present the difference is only of 0.11dex
(see Table~\ref{to/h}). The change is due to two recent results for Orion and
one for the Sun: a) the 0.15dex increase in the O/H value derived from
recombination lines (which implies a $t^2 = 0.024$) relative to that derived
from forbidden lines under the assumption of $t^2 = 0.000$, b) the increase of
0.08dex due to the fraction of oxygen embedded in dust grains, and c) the
decrease of 0.10dex due to a new solar determination.

\begin{table}[htb]
\begin{tabular}{cccc}
\hline
\multicolumn{3}{c}{Orion Nebula} & Sun \\
\cline{1-3}\\[-9pt]
(Gas; $t^2=0.000$) & (Gas; $t^2=0.024$) & ($t^2=0.024$ + Dust) \\
\hline 
$8.49 \pm 0.06^a$ & ... & ... & $8.93 \pm 0.04^b$ \\
$8.47 \pm 0.06^c$ & $8.64 \pm 0.06^c$ & $8.72 \pm 0.07^c$ & $8.83 \pm 0.06^d$\\
\hline

\multicolumn{4}{l}{$^a$ \opencite{sha83}; \opencite{ost92};
\opencite{rub93};}\\
\multicolumn{4}{l}{\hspace*{12pt} \opencite{deh00}.}\\
\multicolumn{4}{l}{$^b$ \opencite{gre89}.}\\
\multicolumn{4}{l}{$^c$ \opencite{est98}.}\\
\multicolumn{4}{l}{$^d$ \opencite{gre98}.}\\
\end{tabular}
\caption{Oxigen abundance for Orion and the Sun (given in log O/H + 12).}
\label{to/h}
\end{table}

To derive the total O/H values in H~{\sc{ii}} regions it is necessary to
estimate the fraction of O embedded in dust grains. For the Orion nebula and
NGC 346 (the brightest H~{\sc{ii}} region in the SMC) it is found that
Fe$_{\rm gas}$/O$_{\rm gas}$ is $1.2 \pm 0.3$dex smaller than in the Sun
(\opencite{est98}; \opencite{rel00}; \opencite{gre98}). For the Orion nebula
and for O poor extragalactic H~{\sc{ii}} regions it is found that Si$_{\rm
gas}$/O$_{\rm gas}$ is $0.46 \pm 0.1$dex and $0.39 \pm 0.1$dex smaller
respectively than in the Sun (\opencite{est98}; \opencite{gar95a};
\opencite{gre98}).  {From} the Si/O and Fe/O underabundances in H~{\sc{ii}}
regions it is estimated that the missing Si and Fe fractions are in dust grains
in the form of molecules that trap about 20\% of the oxygen atoms.

\section{C/O}

The observed C/O ratios are important to test the different sets of stellar
yields present in the literature and the importance of the O-rich galactic
outflows.

\begin{figure}
\centerline{
\includegraphics[bb=0 310 612 720,width=3.3in,clip=true]{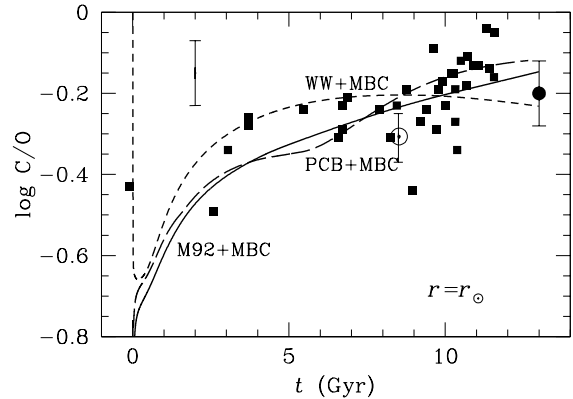}
}
\caption{
C/O evolution of the solar vicinity.
Open circle, solar value from Grevesse \& Sauval (1998). Filled circle,
computed value at $r = r_{\odot}$ from H~{\sc{ii}} region radial gradients by
Peimbert (1999).  Filled squares, dwarf stars at $r = r_{\odot} \pm 1$kpc from
Gustafsson et al. (1999), the ages of the stars were scaled to the age of the
models. Predictions from models assuming yields by Marigo et al. (1996, 1998)
for low and intermediate mass stars in combination with yields for massive
stars by Woosley \& Weaver (1995) and Woosley et al. (1993), Portinari et
al. (1998) or Maeder (1992).  }
\label{f<co-t>}
\end{figure}

The increase of C/O with the age of the disk at the time the stars of the solar
vicinity were formed is due only to the ejecta of massive stars.  Models with
yields by \inlinecite{mae92} or yields by \inlinecite{por98} can reproduce the
increase of C/O with age in the solar neighborhood, while models assuming
yields by \inlinecite{woo95} and \inlinecite{woo93} do not (\opencite{car00};
\opencite{hen00}; \opencite{jin00}).

In Figure~\ref{f<co-t>} we present the evolution of C/O with time for three
different sets of yields, from chemical evolution models by \inlinecite{car00},
as well as the C/O values for a group of dwarf stars of different ages.

{From} chemical evolution models of the Galaxy \inlinecite{car00} finds that
those computations based on yields by \inlinecite{mae92} predict negative C/O
gradients while those based on the yields by \inlinecite{woo95},
\inlinecite{woo93}, or \inlinecite{por98} predict flat gradients. The
observations of negative C/O gradients in our Galaxy (\opencite{pei99} and
references therein), M101 and NGC~2403 \cite{gar99y} support those models based
on the yields by \inlinecite{mae92}.

\begin{table}[htb]
\begin{tabular}{lrlr}
\hline
\multicolumn{1}{c}{Observations} & \multicolumn{1}{c}{$\alpha$} &
\multicolumn{1}{c}{Model Yields} & \multicolumn{1}{c}{$\alpha$} \\
\hline 
Galactic H{~\sc{ii}} regions & $0.93 \pm 0.60^a$ &
\inlinecite{woo95} & $-0.28$ \\
Galactic B stars             & $1.69 \pm 2.34^b$ &
\inlinecite{por98} & $ 0.06$ \\
M101 H{~\sc{ii}} regions     & $1.10 \pm 0.29^c$ &
\inlinecite{mae92} & $ 0.94$ \\
NGC~2403 H{~\sc{ii}} regions & $0.50 \pm 0.43^c$ &
Metal independent  & $ 0.00$ \\
\hline
\multicolumn{4}{l}{$^a$ \citeauthor{est99b},
\citeyear{est98,est99a}, b; \opencite{pei99}.}\\
\multicolumn{4}{l}{$^b$ \opencite{gum98}; \opencite{hib98}.}\\
\multicolumn{4}{l}{$^c$ \opencite{gar99y}.}\\
\end{tabular}
\caption{$\alpha$ values from models of the Galaxy by Carigi (2000) compared
with observations, where $\alpha$ is given by: log C/O = $\alpha$ log O/H.}
\label{talpha}
\end{table}

A powerful way to present the previous result is by means of the parameter
$\alpha$ given by log C/O = $\alpha$ log O/H. In Table~\ref{talpha} we present
the $\alpha$ values for models and observations.

\begin{figure}
\centerline{
\includegraphics[bb=0 310 612 720,width=3.3in,clip=true]{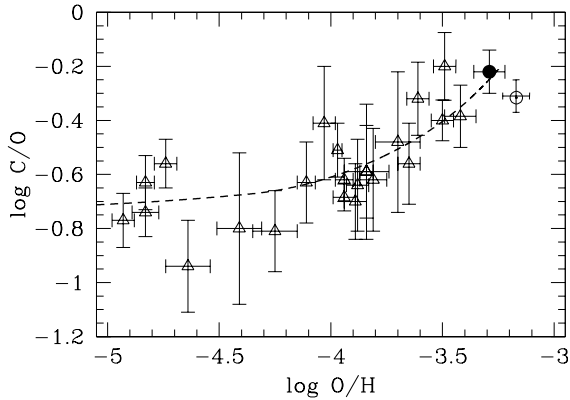}
}
\caption{
Log C/O versus log O/H relation.
Open circle, solar value from Grevesse \& Sauval (1998). Filled circle, Orion
nebula value from Esteban et al. (1998). Open triangles, observational data for
irregular galaxies, M101 and NGC~2403 (Garnett et al. 1995b, 1999; Izotov \&
Thuan 1999). Dashed line, the prediction of the best model for the solar
vicinity by Carigi (2000), which is based on the yields by Maeder (1992) and
van den Hoek \& Groenewegen (1997).  }
\label{f<co-oh>}
\end{figure}

In Figure~\ref{f<co-oh>} we present the best model for the solar vicinity by
\inlinecite{car00} in the C/O versus O/H plane. In this figure we also present
the observed values for the Orion nebula, the Sun, and the extragalactic
H~{\sc{ii}} regions.

\section{$\Delta Y / \Delta O$, $\Delta Y / \Delta Z$}

M17 is the best H{~\sc{ii}} region to determine the helium abundance because
among the brightest Galactic H{~\sc{ii}} regions it is the one with the highest
degree of ionization and consequently with the smallest correction for the
presence of He$^0$(\opencite{pei92}; \opencite{deh00}). By combining the
abundances of M17 and NGC~346 the $\Delta Y / \Delta O$ and $\Delta Y / \Delta
Z$ values presented in Table~\ref{tdelta} were derived, the recommended values
are those for $t^2 = 0.037$.

Based on their two-infall model for the chemical evolution of the Galaxy
\inlinecite{chi97} find $\Delta Y/\Delta O = 3.15$ for the solar
vicinity. \inlinecite{cop97} derives values of $\Delta Y/\Delta O$ in the 2.4
to 3.4 range. \inlinecite{car00} computed chemical evolution models for the
Galactic disk, under an inside-out formation scenario, based on different
combinations of seven sets of stellar yields by different authors; the $\Delta
Y/\Delta O$ spread predicted by her models is in the 2.9 to 4.6 range for the
Galactocentric distance of M17 (5.9~kpc), the spread is only due to the use of
different stellar yields.  For massive stars $\Delta Y/\Delta O$ increases
along the sequence \inlinecite{por98} $\to$ \inlinecite{mae92} $\to$
\inlinecite{woo95}, while for intermediate mass stars it increases along the
sequence \inlinecite{hoe97} $\to$ \inlinecite{ren81} $\to$
\inlinecite{mar96}. The differences between all the models and the observations
for $t^2 = 0.000$ are significant, while the differences between some of the
models and the observations for $t^2 = 0.037$ probably are not.

{From} a group of 10 irregular and blue compact galaxies \inlinecite{car95}
found $\Delta Y/\Delta O = 4.48 \pm 1.02$, where they added 0.2~dex to the O/H
abundance ratios derived from the nebular data to take into account the
temperature structure of the H{~\sc{ii}} regions and the fraction of O embedded
in dust; moreover they also estimated that $O$ constitutes 54\% of the $Z$
value. \inlinecite{izo98} from a group of 45 supergiant H{~\sc{ii}} regions of
low metalicity derived that $\Delta Y/\Delta Z = 2.3 \pm 1.0$; we find from
their data that $\Delta Y/\Delta Z = 1.46 \pm 0.60$ by adding 0.2~dex to the O
abundances to take into account the temperature structure of the H{~\sc{ii}}
regions and the fraction of O embedded in dust; furthermore from their data we
also find that $\Delta Y/\Delta O = 2.7 \pm 1.2$ by assuming that $O$
constitutes 54\% of the $Z$ value.

\inlinecite{car95}, based on yields by \inlinecite{mae92}, computed closed box
models adequate for irregular galaxies obtaining $\Delta
Y/\Delta O = 2.95$. They also computed models with galactic outflows of well
mixed material that yielded $\Delta Y/\Delta O$ values similar to those of the
closed box models, and models with galactic outflows of O-rich material that
yielded values higher than 2.95. The maximum $\Delta Y/\Delta O$ value that can
be obtained with models of O-rich outflows, without entering into contradiction
with the C/O and $(Z-C-O)/O$ observational constraints, amounts to 3.5.

\inlinecite{car99}, based on yields by \inlinecite{woo93} and
\inlinecite{woo95}, computed chemical evolution models for irregular galaxies
and found very similar values for closed box models with bursting star
formation and constant star formation rates that amounted to $\Delta Y/\Delta O
= 4.2$. The models with O-rich outflows can increase the $\Delta Y/\Delta O$,
but they predict higher C/O ratios than observed.

O-rich outflows are not very important for the typical irregular galaxy because
they predict C/O and $Z/O$ ratios higher than observed. \inlinecite{lar00}
reach the same conclusion based on models to explain the N/O ratios.

\begin{table}[htb]
\begin{tabular}{lrc}
\hline
\multicolumn{1}{c}{Object} &
\multicolumn{1}{c}{$\Delta Y/\Delta O$} &
\multicolumn{1}{c}{$\Delta Y/\Delta Z$} \\
\hline 
M17 $(t^2=0.000)^a$                      & $13.3 \pm 2.7$ & $ 3.8 \pm 1.1$ \\
M17 $(t^2=0.037)^a$                      &  $5.4 \pm 1.1$ & $ 2.1 \pm 0.6$ \\
Solar vicinity models$^b$                &   2.4  -- 4.6  &   1.1  -- 2.1  \\
Irregular galaxies, observations$^{c,d}$ &  $3.5 \pm 1.1$ & $ 1.9 \pm 0.6$ \\
Irregular galaxies, models$^{c,e}$       &   2.9  -- 4.2  &   1.6  -- 2.3  \\
\hline

\multicolumn{3}{l}{$^a$ \citeauthor{pei92}, \citeyear{pei92,pei00a};
\opencite{est99a}.}\\
\multicolumn{3}{l}{$^b$ \opencite{cop97}; \opencite{chi97};
\opencite{car00}.}\\
\multicolumn{3}{l}{$^c$ \opencite{car95}.}\\
\multicolumn{3}{l}{$^d$ \opencite{izo98}.}\\
\multicolumn{3}{l}{$^e$ \opencite{car99}.}\\
\end{tabular}
\caption{Helium to oxygen and helium to heavy element ratios by mass:
$\Delta Y/\Delta O$ and $\Delta Y/\Delta Z$.}
\label{tdelta}
\end{table}

\section{Primordial Helium Abundance, $Y_p$}

Recent discussions on the determination of $Y_p$ have been presented by
\inlinecite{thu00} and \inlinecite{pei00z}. \inlinecite{izo98}, from the $Y$ --
O/H linear regression for a sample of 45 BCGs, and \inlinecite{izo99z}, from
the average for the two most metal deficient galaxies known (I~Zw~18 and
SBS~0335--052), derive $Y_p$ values of $0.2443\pm 0.0015$ and $0.2452 \pm
0.0015$ respectively. Alternatively, \citeauthor{pei00b} (\citeyear{pei00a},
b), based on NGC~346, NGC~2363, and I~Zw~18, derive $Y_p = 0.2351 \pm 0.0022$.
Most of the difference is due to the $T_e$(He~{\sc{ii}}) used by both groups,
while \citeauthor{izo98} and \citeauthor{izo99z} assume that
$T_e$(He~{\sc{ii}}) equals $T_e$(O~{\sc{iii}}), \citeauthor{pei00a} find that
$T_e$(He~{\sc{ii}}) is about 9\% smaller than $T_e$(O~{\sc{iii}}).

Under the framework of standard Big Bang nucleosynthesis computations it is
possible to compare the $Y_p$, D$_p$, and Li$_p$ values through the predicted
$\Omega_b$ values.

The high $Y_p$ determination of $0.2452 \pm 0.0015 (1\sigma)$ combined with
standard Big Bang nucleosynthesis computations (\opencite{tho94};
\opencite{fio98}) implies that, at the $1\sigma$ confidence level, $\Omega_b
h^2$ is in the 0.0139 to 0.0190 range.  For $h = 0.65$ the $Y_p$ value
corresponds to $0.033 < \Omega_b < 0.045$, a value in very good agreement with
that derived from the primordial deuterium abundance, D$_p$, determined by
\inlinecite{bur98} that amounts to $0.041 < \Omega_b < 0.047 (1\sigma)$ for $h
= 0.65$.

The low $Y_p$ determination of $0.2351 \pm 0.0022 (1\sigma)$ implies that, at
the $1\sigma$ confidence level, $\Omega_b h^2$ is in the 0.0060 to 0.0081
range.  For $h = 0.65$ the $Y_p$ value corresponds to $0.014 < \Omega_b <
0.019$, a value in good agreement with that derived from the primordial lithium
abundance, Li$_p$, determined by \inlinecite{suz00} that amounts to $0.015 <
\Omega_b < 0.033 (2\sigma)$ for $h = 0.65$, in very good agreement with the low
redshift estimate of the global budget of baryons by \inlinecite{fuk98} who
find $0.015 < \Omega_b < 0.030 (1\sigma)$ for $h = 0.65$, and consistent with
their minimum to maximum range for redshift $z = 3$ that amounts to $0.012 <
\Omega_b < 0.070$ for $h = 0.65$.

The discrepancy between the low $Y_p$ value and the D$_p$ value should be
studied further.

\theendnotes

\end{article}
\end{document}